\documentclass[12pt]{article}

\usepackage[dvips]{graphicx}
\makeatletter
 
 \@addtoreset{equation}{section}
\makeatother 
\newcommand{\sumtwo}[2]{\mathop{\sum_{#1}}_{#2}}

\newtheorem{proposition}{Proposition}[section]

\newcommand{\calK}{\mathcal{K}}
\newcommand{\calB}{\mathcal{B}}

\newcommand{\La}{\Lambda}
\newcommand{\up}{\uparrow}
\newcommand{\dn}{\downarrow}

\newcommand{\sgn}{{\mathbf{sgn}}}
\newcommand{\rme}{\mathrm{e}}
\newcommand{\rmi}{\mathrm{i}}
\newcommand{\Ne}{{N_\mathrm{e}}}
\newcommand{\Nh}{{N_\mathrm{h}}}

\newcommand{\vecS}{\mbox{\boldmath $S$}}
\newcommand{\PhiG}{\Phi_{\mathrm{G}}}
\newcommand{\PhiF}{\Phi_{\mathrm{F}}}
\newcommand{\vecd}{\mbox{\boldmath $d$}}
\newcommand{\bsigma}{\mbox{\boldmath $\sigma$}}
\newcommand{\btau}{\mbox{\boldmath $\tau$}}
\newcommand{\bdelta}{\mbox{\boldmath $\delta$}}
\newcommand{\ba}{\mbox{\boldmath $a$}}
\newcommand{\bx}{{\mbox{\boldmath $x$}}}
\newcommand{\by}{{\mbox{\boldmath $y$}}}

\newcommand{\sba}{{\mbox{\scriptsize\boldmath $a$}}}
\newcommand{\sbx}{{\mbox{\scriptsize\boldmath $x$}}}
\newcommand{\sby}{{\mbox{\scriptsize\boldmath $y$}}}
\newcommand{\sbk}{{\mbox{\scriptsize\boldmath $k$}}}
\newcommand{\bvarphi}{\mbox{\boldmath $\varphi$}}
\newcommand{\btvarphi}{\tilde{\mbox{\boldmath ${\varphi}$}}}
\newcommand{\Hhop}{H_\mathrm{hop}}
\newcommand{\HintJ}{H_{\mathrm{int},J}}
\newcommand{\tHintJ}{\bar{H}_{\mathrm{int},J}}
\newcommand{\HintU}{H_{\mathrm{int},U}}
\newcommand{\cxs}{c_{x,\sigma}}
\newcommand{\cxsd}{c_{x,\sigma}^\dagger}
\newcommand{\bxs}{b_{x,\sigma}}
\newcommand{\bxsd}{b_{x,\sigma}^\dagger}
\newcommand{\tbxs}{\tilde{b}_{x,\sigma}}
\newcommand{\tbxsd}{\tilde{b}_{x,\sigma}^\dagger}

\newcommand{\eqref}[1]{(\ref{#1})}
\begin{document}
\begin{center}
\textbf{\large 
Ferromagnetic Pairing States on Two-Coupled Chains
}\bigskip\\
Akinori Tanaka%
\footnote{
 akinori@ariake-nct.ac.jp
}\bigskip\\
\textit{Department of General Education, 
          Ariake National College of Technology,
        Omuta 836-8585, Japan}\bigskip\\
(March 3, 2008)
\end{center}
\vspace*{4cm}
\begin{abstract}
We propose a concrete model which exhibits ferromagnetism and
electron-pair condensation simultaneously.
The model is defined on two chains and 
consists of the electron hopping term,   
the on-site Coulomb repulsion, and a ferromagnetic interaction 
which describes ferromagnetic coupling between two electrons,
one on a bond in a chain and the other on a site in the other chain.
It is rigorously shown that the model has fullypolarized ferromagnetic
 pairing ground states.
The higher dimensional version of the model is also presented.
\end{abstract}
\newpage
\section{Introduction}
Recently, UGe$_2$~\cite{Saxena00}, URhGe~\cite{Aoki01}, and
UCoGe~\cite{Huy07} 
were discovered to exhibit
ferromagnetic superconductivity.
Experimental results suggest that the same electrons are
contributing to both ferromagnetism and superconductivity,
and thus, in the superconducting phase of these materials, the electrons are considered
to condensate into a spin-triplet pair state unlike usual low temperature 
superconductors, in which electrons are 
forming non-magnetic spin-singlet pairs.
Microscopic explanation of this phenomenon is a challenge in condensed
matter physics, but the problem is rather subtle and difficult
since we have to treat spin-rotation symmetry breaking 
and electron-pair condensation simultaneously.
In fact, 
the mechanism for ferromagnetism alone in itinerant electron systems
has not yet been fully understood,
although there have been some rigorous 
developments~\cite{Mielke91,MielkeTasaki93,Tasaki98,Tasaki03,Tanaka03,TanakaTasaki07}
in the Hubbard model, which is one of the simplest models
of itinerant electron systems.
As for electron-pair condensation, some models have been shown to have pairing
ground
states~\cite{Essler92,deBoer95,Tanaka04,Tanaka04-2,TanakaYamanaka05,Tanaka07},
but little is known about spin-triplet pairing ground states.

The Coulomb repulsion between electrons combined with the Pauli exclusion principle
can generate ferromagnetism 
in itinerant electron systems~\cite{Mielke91,MielkeTasaki93,Tasaki98,Tasaki03,Tanaka03,TanakaTasaki07}, 
but it is unclear at the present time that the same can also lead to
ferromagnetic superconductivity.
Since the electrons form spin-triplet pairs, it is expected that
there appears a kind of effective ferromagnetic interaction between
electrons in the ferromagnetic superconductors.
Thus a model containing ferromagnetic interactions 
as well as the Coulomb repulsion between
electrons will be suitable for an investigation 
as a first step toward understanding microscopic mechanisms 
for ferromagnetic superconductivity.

\begin{figure}
 \label{fig:image}
 \begin{center}
   \includegraphics[width=.5\textwidth]{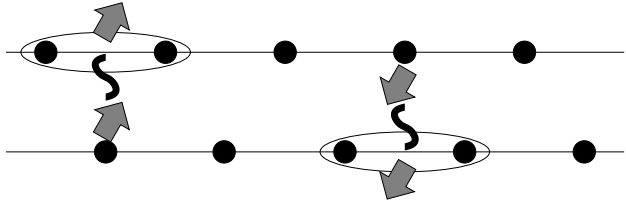}
 \end{center}
 \caption{The image of the ferromagnetic interaction considered
 here. The solid circles represent lattice sites, and thick arrows
 represent electron spins. The electron on a bond in a chain favors to
 couple ferromagnetically with the one on a site in the other chain.}
\end{figure}

In this paper we propose a model which
has fullypolarized ferromagnetic pairing ground states.
The model consists of
the electron hopping term, the on-site Coulomb repulsion 
and a short-range ferromagnetic interaction term.
It is possible to consider the model in any dimension, but 
for simplicity
we mainly concentrate
on a one-dimensional version here.
In the one-dimensional case, our model is defined on two chains.
The electrons can hop along the chain direction, feeling on-site repulsion,
and furthermore the two electrons, one on a bond in a chain and the
other on a site in the other chain, feel a ferromagnetic interaction
(see Fig.~\ref{fig:image}).
It is shown in our model that, 
owing to the ferromagnetic interaction introduced here, 
the electrons form spin-triplet pairs, 
and then the fullypolarized ferromagnetic
pairing state is selected by the on-site Coulomb repulsion  
as the unique ground state (up to degeneracy due to
spin-rotation symmetry).
It is also shown that the ground state exhibits off-diagonal long-range
order associated with local spin-triplet pairs. 

This paper is organized as follows.
In the next section we set up the model 
and state the main results.
In section~\ref{s:proof} we prove the main results.
In section~\ref{s:remarks} we make some remarks on our model and
ferromagnetic pairing states.
We also present models with anisotropic spin-interactions 
and show that the models exhibit other kinds of spin-1 electron-pair condensation.  
In the final section we briefly comment on the higher dimensional
version of the model.

\section{Definition and Main Results}
\label{s:definition and main results}
We start by defining a lattice of our model.
Let $\La_1$ and $\La_2$ be linear chains with $L$ sites.
We label sites in $\La_1$ and $\La_2$ by integers, and half integers,
respectively, as $\La_1=\{0,1,\dots,L-1\}$ and
$\La_2=\{1/2,3/2,\dots,L-1/2\}$.
It is assumed that $L$ is a positive odd integer, and periodic boundary
conditions are imposed. 
Then let $\La=\La_1\cup\La_2$, on which our Hamiltonian will be
defined.

We next introduce fermion operators.
As usual, we denote by $\cxs$ and $\cxsd$ the annihilation and creation
operators, respectively, of an electron with spin $\sigma=\up,\dn$ 
at site $x\in\La$. 
They satisfy the canonical anticommutation relations, 
\begin{equation}
\label{eq:c-operator-anticommute1}
 \{c_{x,\sigma},~c_{y,\tau}\}=
  \{c_{x,\sigma}^\dagger,~c_{y,\tau}^\dagger\}=0,
\end{equation}
and
\begin{equation}
\label{eq:c-operator-anticommute2}
 \{c_{x,\sigma}^\dagger,~c_{y,\tau}\}=\delta_{x,y}\delta_{\sigma,\tau}
\end{equation}
for any $x,y\in\La$ and $\sigma,\tau=\up,\dn$.
We denote by $\Ne$ the electron number and by $\Phi_0$ a state without
electrons.
An $\Ne$-electron state can be constructed by operating 
$\Ne$ creation operators $c_{x,\sigma}^\dagger$ on $\Phi_0$.

\begin{figure}
\label{fig:components}
 \begin{center}
  \begin{tabular}{ccc}
    \includegraphics[width=.45\textwidth]{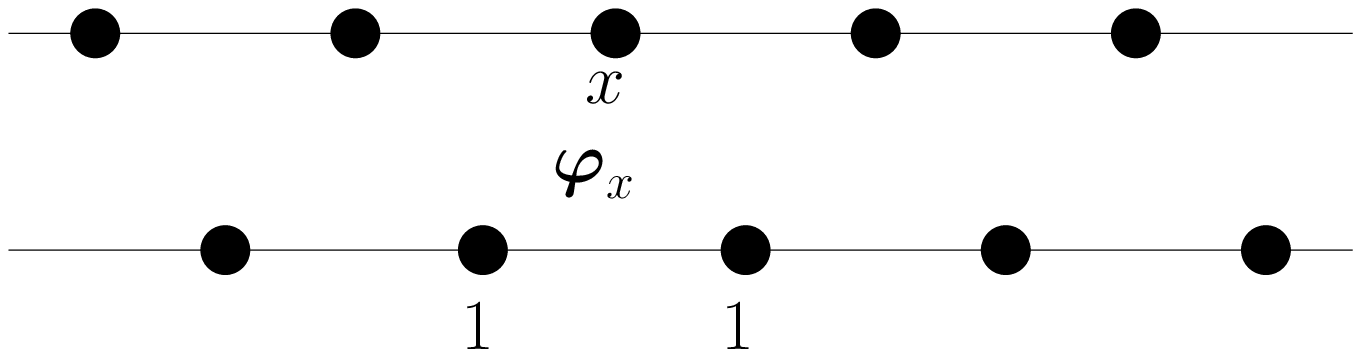}
     & \hspace*{.02\textwidth}
    \includegraphics[width=.45\textwidth]{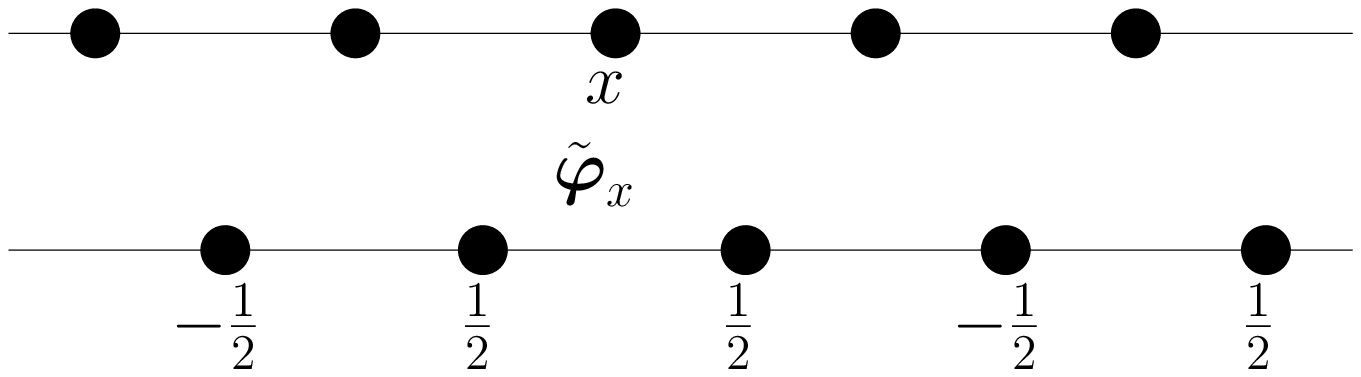}
  \end{tabular}
 \end{center}
 \caption{The components of the vectors $\bvarphi_x$ and $\btvarphi_x$.}
\end{figure}

We also introduce other fermion operators
which play an essential role in our model.
For each $x\in\La$, let $\bvarphi_x=(\varphi_x(y))_{y\in\La}$ 
be a vector whose components are given by~\cite{comment1}
\begin{equation}
 \varphi_x(y)=\left\{
	       \begin{array}{@{\,}l}
		     1 ~~~\mbox{if $|y-x|=1/2$},\\
                     0 ~~~\mbox{otherwise}   
	       \end{array}   
              \right.,
\end{equation}
(see Fig.~\ref{fig:components}) and define 
\begin{equation}
 \bxs=\sum_{y\in\La}\varphi_x(y)c_{y,\sigma}=c_{x-\frac{1}{2},\sigma}+c_{x+\frac{1}{2},\sigma}.
\label{eq:b}
\end{equation}
This operator corresponds to a single-electron
state localized at a bond of nearest neighbour sites in a chain.
Since the set of all vectors $\bvarphi_x$ is not orthonormal,
the $b$-operators do not satisfy the canonical anticommutation relations 
(in fact, it is easy to see that
$\{b_{x,\sigma}^\dagger,~b_{y,\sigma}\}=1$ if $|x-y|=1$),
so that we introduce dual operators $\tbxs$ as follows.
First recall that we adopt the
periodic boundary conditions, and thus, 
for each $x\in\La_1$, sites $y\in\La_2$ can be written as
$y=x+1/2+n(x,y)$ where $n(x,y)$ is an integer in $0\le n(x,y) \le L-1$.
Then, for $x\in\La_1$, let
$\btvarphi_x=(\tilde{\varphi}_{x}(y))_{y\in\La}$ 
be a vector whose components are given by  
\begin{equation}
 \tilde{\varphi}_x(y)=\left\{
	       \begin{array}{@{\,}rl}
		     1/2 &~~~\mbox{if $y\in\La_2$ and $ n(x,y)$ is even}\\
                     -1/2 &~~~\mbox{if $y\in\La_2$ and $n(x,y)$ is odd} \\   
                     0 &~~~\mbox{if $y\in\La_1$ },   
	       \end{array}   
              \right.
\end{equation}
(see Fig.~\ref{fig:components}) and, for $x\in\La_2$, let $\btvarphi_x$ be a vector obtained by 
$\tilde{\varphi}_{x}(y)=\tilde{\varphi}_{x-1/2}(y-1/2)$.
It is easy to see that the inner product of $\btvarphi_x$ and $\bvarphi_y$ satisfies
\begin{equation}
 \langle \btvarphi_x,\bvarphi_y \rangle
  = \sum_{z\in\La}\tilde{\varphi}_x^\ast(z)\varphi_y(z)
  = \delta_{x,y}.
\end{equation}
By using $\btvarphi_x$, we define
\begin{equation}
\label{eq:tildeb}
 \tbxs = \sum_{y\in\La}\tilde{\varphi}_x(y)c_{y,\sigma}.
\end{equation}
The $\tilde{b}$-operators are dual to the $b$-operators
in the sense 
\begin{equation}
\{\tbxs^\dagger, b_{y,\tau}\}
=\langle \btvarphi_x,\bvarphi_y \rangle\delta_{\sigma,\tau}
=\delta_{x,y}\delta_{\sigma,\tau}.
\label{eq:tildeb-b}
\end{equation}
The anticommutation relation~\eqref{eq:tildeb-b} implies that 
the set $\{\bxsd\Phi_0\}_{x\in\La}$ is linearly independent 
and so is the set $\{\tbxsd\Phi_0\}_{x\in\La}$.
Furthermore, \eqref{eq:tildeb-b} implies that the $c$-operators 
can be represented in terms of the $\tilde{b}$-operators as
\begin{eqnarray}
  \cxs&=&\sum_{y\in\La}\{b_{y,\sigma}^\dagger,~\cxs\}\tilde{b}_{y,\sigma}
      =\sum_{y\in\La}\varphi_y^\ast(x)\tilde{b}_{y,\sigma}
\nonumber\\
      &=& \sum_{y\in\La}\varphi_x(y)\tilde{b}_{y,\sigma}
       =\tilde{b}_{x-\frac{1}{2},\sigma}+\tilde{b}_{x+\frac{1}{2},\sigma}.
\label{eq:inverse-tildeb}
\end{eqnarray}  

Let us define the number operators $n_{x,\sigma}$ with $\sigma=\up,\dn$
and the spin operators $S_{x,\alpha}$ with $\alpha=1,2,3$
for the $c$-operators by
\begin{eqnarray}
\label{eq:number-operator1}
 n_{x,\sigma}&=&\cxsd\cxs, \\
 S_{x,1}     &=&\frac{1}{2}\left(c_{x,\up}^\dagger c_{x,\dn} 
                               + c_{x,\dn}^\dagger c_{x,\up}\right), \\
 S_{x,2}     &=&\frac{1}{2\rmi}\left(c_{x,\up}^\dagger c_{x,\dn} 
                               - c_{x,\dn}^\dagger c_{x,\up}\right), \\
\label{eq:spin-operator}
 S_{x,3}     &=&\frac{1}{2}\left(c_{x,\up}^\dagger c_{x,\up} 
                               - c_{x,\dn}^\dagger c_{x,\dn}\right). 
\end{eqnarray}
We also define 
\begin{equation}
\label{eq:number-operator2}
n_x=n_{x,\up}+n_{x,\dn}.
\end{equation}
The number operators $n_{x,\sigma}^b$ and $n_x^b$, and the spin
operators $S_{x,\alpha}^b$ for the $b$-operators 
are defined similarly by using $b_{x,\sigma}$ in place of $c_{x,\sigma}$
in \eqref{eq:number-operator1} -- \eqref{eq:number-operator2}.

By using the operators introduced as above, we define the Hamiltonian $H$ as
follows:
\begin{eqnarray}
 \Hhop &=& t\sum_{x\in\La}\sum_{\sigma=\up\dn}b_{x,\sigma}^\dagger b_{x,\sigma},\\
 \HintJ &=& -J\sum_{x\in\La}
                 \left(\frac{n_{x}^b n_{x}}{4}+\vecS_{x}^b\cdot\vecS_x\right),\\
 \HintU &=& U\sum_{x\in\La} n_{x,\up}n_{x,\dn},\\
     H  &=& \Hhop+\HintJ+\HintU,
\end{eqnarray} 
where $t,J>0$ and $U\ge0$.
By using the $c$-operators, $\Hhop$ is rewritten as  
\begin{equation}
 \Hhop = t\sumtwo{x,y\in\La}{|x-y|=1}\sum_{\sigma=\up,\dn} 
                       c_{x,\sigma}^\dagger c_{y,\sigma}+2t\sum_{x\in\La}n_x,
\end{equation}
which represents the motion of electrons in the two chains.
The Hamiltonian $\HintU$ represents the repulsive 
interaction between two electrons with
the opposite spins at the same site 
and $\Hhop+\HintU$ reduces to the usual Hubbard Hamiltonian.
In our model, owing to the Hamiltonian $\HintJ$,    
an electron at a bond in one chain and an electron at a site in
the other chain feel the attractive and ferromagnetic interaction.
We stress that all the interaction terms considered here are of short-range.
It is also noted that the Hamiltonian $H$ conserves the electron number
and possesses spin-rotation symmetry.
The occurrence of ferromagnetism and electron-pair condensation 
in our model is thus not trivial at all.  

Let $[G_{x,y}]_{x,y\in\La}$ be an antisymmetric matrix whose elements
are given by
\begin{equation}
 G_{x,y}= \left\{
	       \begin{array}{@{\,}rl}
		     -1/2  &~~~\mbox{if $x\in\La_1$, $y\in\La_2$ and $|x-y|=1/2$}\\
                     1/2 &~~~\mbox{if $x\in\La_2$, $y\in\La_1$ and $|x-y|=1/2$} \\   
                     0  &~~~\mbox{otherwise},   
	       \end{array}   
              \right.
\end{equation}
and define $\tilde{\zeta}_{\sigma\tau}$ and $\zeta_{\sigma\tau}$,
corresponding to pair states with spin-1, by
\begin{equation}
 \tilde{\zeta}_{\sigma\tau}=\sum_{x,y\in\La}G_{x,y}\tilde{b}_{x,\sigma}\tilde{b}_{y,\tau}
\end{equation}
and 
\begin{equation}
 {\zeta}_{\sigma\tau}=\sum_{x,y\in\La}G_{x,y}{c}_{x,\sigma}{c}_{y,\tau},
\end{equation}
respectively.

Our main results are summarized as follows:
{\proposition
\label{prop:ground-state}
Assume that the electron number $\Ne$ is an even integer in $2\le\Ne\le|\La|$,
and consider the Hamiltonian $H$ with $2t=J$ and $U>0$. 
Then the fullypolarized ferromagnetic pairing state,
\begin{equation}
 \PhiG=\left(\tilde{\zeta}_{\up\up}^\dagger\right)^{\frac{\Ne}{2}}\Phi_0,
\end{equation}
is the ground state of $H$.
Furthermore the ground state is unique 
up to degeneracy due to spin-rotation symmetry.
}
{\proposition
\label{prop:order}
Consider the Hamiltonian $H$ with $2t=J$ and $U>0$, and take a sequence of the
ground states $\PhiG$ of $H$ for even $\Ne$ and $\La$ such that 
the electron filling factor $\Ne/(2|\La|)$ converges to $\nu$ in $0<\nu\le1/2$.
Let $\Delta=\zeta_{\up\up}/L$ and $\displaystyle g_k=2\cos(k/2)$.
Then we have
\begin{equation}
 \mu(\nu)=\lim_{|\La|,\Ne\to\infty}
               \frac{\langle \PhiG,\Delta \Delta^\dagger  \PhiG\rangle}
              {\langle \PhiG,\PhiG\rangle}
         \ge \frac{1}{2}\left(1-2\nu\right)I(\nu)
\label{eq:order}
\end{equation} 
with
\begin{equation}
 I(\nu)=2\left(
           \frac{1}{2\pi}\int_{|k|\le \pi} \chi[|g_k|^2\le \epsilon(\nu)]|g_k|^2 \mathrm{d}k
               \right),
\end{equation}
where $\chi[X]$ equals $1$ if $X$ is true and zero otherwise, and
${\epsilon}(\nu)$ is determined by
\begin{equation}
 \nu = \frac{1}{2}\left(
	       \frac{1}{2\pi}\int_{|k|\le \pi} \chi[|g_k|^2\le \epsilon(\nu)] \mathrm{d}k
              \right).
\end{equation}
}
\section{Proof}
\label{s:proof}
\textbf{Proof of Proposition \ref{prop:ground-state}}.
In the proof we fix the electron number $\Ne$ to an even integer 
in $2\le\Ne\le|\La|$. 

By some straightforward calculations, we can rewrite $\HintJ$ as
\begin{equation}
 \HintJ=-\frac{J}{2}\sum_{x\in\La}
                         (b_{x\up}^\dagger b_{x,\up}c_{x,\up}^\dagger c_{x,\up}
                         +b_{x\dn}^\dagger b_{x,\dn}c_{x,\dn}^\dagger c_{x,\dn}
                         +b_{x\up}^\dagger b_{x,\dn}c_{x,\dn}^\dagger c_{x,\up}
                         +b_{x\dn}^\dagger b_{x,\up}c_{x,\up}^\dagger c_{x,\dn}).
\end{equation}
Noting the assumption $2t=J$ and the anticommutation relations 
\eqref{eq:c-operator-anticommute1},
\eqref{eq:c-operator-anticommute2} and $\{\bxsd,\cxs\}=0$, 
we obtain
\begin{equation}
\label{eq:Hamiltonian2}
 H=\tHintJ+\HintU
\end{equation}
with
\begin{equation}
 \tHintJ=\frac{J}{2}\sum_{x\in\La}(b_{x,\up}^\dagger c_{x,\up} + b_{x,\dn}^\dagger c_{x,\dn})
                             (c_{x,\up}^\dagger b_{x,\up} + c_{x,\dn}^\dagger b_{x,\dn}). 
\end{equation}

Since
$\tHintJ$ and $\HintU$ are sums of positive semi-definite operators 
\begin{equation}
\frac{J}{2}(b_{x,\up}^\dagger c_{x,\up} + b_{x,\dn}^\dagger c_{x,\dn})
                             (c_{x,\up}^\dagger b_{x,\up} +
			     c_{x,\dn}^\dagger b_{x,\dn})
\end{equation}
and 
\begin{equation}
Un_{x,\up}n_{x,\dn}=Uc_{x,\up}^\dagger c_{x,\dn}^\dagger c_{x,\dn}c_{x,\up},
\end{equation}
respectively,
the energy eigenvalues of $\tHintJ+\HintU$ are
greater than or equal to zero. 
Therefore, one can conclude that a zero-energy state $\Phi$ of both of these
Hamiltonians, which satisfies
\begin{equation}
 (c_{x,\up}^\dagger b_{x,\up} +
			     c_{x,\dn}^\dagger b_{x,\dn})\Phi=0
\label{eq:condition-J}
\end{equation}
and 
\begin{equation}
 c_{x,\dn}c_{x,\up}\Phi=0
\label{eq:condition-U}
\end{equation} 
for all $x\in\La$
(if it exists), is a ground state. 
We shall show that the state $\PhiG$ is indeed a zero-energy state of 
both $\tHintJ$ and  $\HintU$, and thus is a ground state of $H$.

Since there is no electron with down-spin in $\PhiG$, the operation of
$c_{x,\up}^\dagger b_{x,\up} + c_{x,\dn}^\dagger b_{x,\dn}$ on $\PhiG$
reduces to $c_{x,\up}^\dagger b_{x,\up}\PhiG$.
Here, by using \eqref{eq:tildeb-b} and \eqref{eq:inverse-tildeb}, we have 
\begin{equation}
 c_{x,\up}^\dagger b_{x,\up}\tilde{\zeta}_{\up\up}^\dagger
            =c_{x,\up}^\dagger \left\{(-1)^{l+1}c_{x,\up}^\dagger  
                                  +\tilde{\zeta}_{\up\up}^\dagger b_{x,\up}\right\}
            =\tilde{\zeta}_{\up\up}^\dagger c_{x,\up}^\dagger b_{x,\up}
\label{eq:tzeta-commutation}
\end{equation}
for $x\in\La_l$ with $l=1,2$. Therefore, we have 
$c_{x,\up}^\dagger b_{x,\up}\PhiG=0$,
from which $\tHintJ\PhiG=0$ follows.  
Furthermore,
noting again that there is no electron with down-spin in $\PhiG$, one
immediately finds $\HintU\PhiG=0$.
We conclude that $\PhiG$ is a zero-energy state of $\tHintJ+\HintU$.   

In what follows, we shall show that any zero-energy state of
$\tHintJ+\HintU$ must be expanded in terms of $\PhiG$
and its $SU(2)$ rotations.

Assume that $\Phi$ is a zero-energy state of $H$ 
and thus satisfies \eqref{eq:condition-J} and
\eqref{eq:condition-U} for all $x\in\La$.
We firstly show that $\Phi$ can be written as~\cite{comment2}
\begin{equation}
\label{eq:Phi-expansion0}
 \Phi=\sumtwo{C\subset\La_1}{|C|=\Ne/2}\sum_{\bsigma,\btau}f(C;\bsigma,\btau)
                    \left(\prod_{x\in C}\tilde{b}_{x,\sigma_x}^\dagger\right)
                    \left(\prod_{x\in C}c_{x,\tau_x}^\dagger\right)
                    \Phi_0,
\end{equation}
with coefficients $f$, where $\bsigma$ and $\btau$ are short hands of spin configurations
$(\sigma_{x})_{x\in C}$ and $(\tau_x)_{x\in C}$ with
$\sigma_x,\tau_x=\up,\dn$, respectively,
and the sum
$\sum_{\bsigma,\btau}$ runs over all possible spin configurations. 

To prove the above claim, we begin with preparing basis states for
$\Ne$-electron states.
Noting the linear independence of $\tilde{b}$-operators and the
relation~\eqref{eq:inverse-tildeb}, 
we find that all states of the form
\begin{equation}
\label{eq:Phi2}
 \Phi_2(B_\up,B_\dn;C_\up,C_\dn)
 =
  \left(\prod_{x\in B_\up}\tilde{b}_{x,\up}^\dagger\right)
  \left(\prod_{x\in B_\dn}\tilde{b}_{x,\dn}^\dagger\right)
  \left(\prod_{x\in C_\up}c_{x,\up}^\dagger\right)
  \left(\prod_{x\in C_\dn}c_{x,\dn}^\dagger\right)
  \Phi_0
\end{equation}
with subsets $B_\sigma$, $C_\sigma$  of $\La_2$ such that
$\sum_{\sigma=\up,\dn}(|B_{\sigma}|+|C_{\sigma}|)=\Ne$
form complete basis states for the $\Ne$-electron Hilbert space.
By using these basis states  we expand $\Phi$ as
\begin{equation}
 \Phi=\sum_{B_\up,B_\dn\subset\La_2}\sum_{C_\up,C_\dn\subset\La_2}
           f_2(B_\up,B_\dn;C_\up,C_\dn)\Phi_2(B_\up,B_\dn;C_\up,C_\dn),
\end{equation} 
with coefficients $f_2$, where the sum runs over all possible subsets of $\La_2$ such that 
$\sum_{\sigma=\up,\dn}(|B_{\sigma}|+|C_{\sigma}|)=\Ne$.
(For notational simplicity, here and in the rest of the proof we do not explicitly write
this restriction in the expressions.)

Let us derive the condition on $f_2$ 
which follows from~\eqref{eq:condition-J} and \eqref{eq:condition-U}.
We choose a site $z\in\La_2$ and $B_\sigma^\prime,C_\sigma\subset\La_2$
with $\sigma=\up,\dn$ 
satisfying $z\notin B_\sigma^\prime$ and
$\sum_{\sigma}(|B_\sigma|+|C_\sigma|)=\Ne-2$.
Then, from \eqref{eq:condition-J}, we have that
\begin{equation}
  \left(\prod_{x\in B_\dn^\prime}{b}_{x,\dn}\right)
  \left(\prod_{x\in B_\up^\prime}{b}_{x,\up}\right)
  \left(\prod_{x\in C_\dn}c_{x,\dn}\right)
  \left(\prod_{x\in C_\up}c_{x,\up}\right) 
    b_{z,-\sigma}c_{z,\sigma}(c_{z,\up}^\dagger b_{z,\up} +
			     c_{z,\dn}^\dagger b_{z,\dn})\Phi=0,
\end{equation}
which leads to
$f_2(B_\up^\prime\cup\{z\},B_\dn^\prime\cup\{z\};C_\up,C_\dn)=0$ 
for $z\notin C_\sigma$.
Therefore,
if $z\notin C_\up\cap C_\dn$,
then
$f_2(B_\up^\prime\cup\{z\},B_\dn^\prime\cup\{z\};C_\up,C_\dn)=0$ 
or equivalently 
$f_2(B_\up,B_\dn;C_\up,C_\dn)=0$ 
for $B_\up$ and $B_\dn$ such that $z\in B_\up\cap B_\dn$.
On the other hand, by examining \eqref{eq:condition-U} for sites in $\La_2$, 
we also obtain $f_2(B_\up,B_\dn;C_\up,C_\dn)=0$ for 
$C_\up$ and $C_\dn$ with $C_\up\cap C_\dn\ne\emptyset$.
As a result we find 
that $f_2(B_\up,B_\dn;C_\up,C_\dn)=0$  
unless $B_\up\cap B_\dn=\emptyset$ as well as $C_\up\cap
C_\dn=\emptyset$ is satisfied.
 
We furthermore examine~\eqref{eq:condition-J}.  
Fix $C_\up$ and $C_\dn$ with $C_\up\cup C_\dn\ne\La_2$ 
and choose a site $z\in\La_2$ which is not
included in $C_\up\cup C_\dn$.
Using the condition~\eqref{eq:condition-J} for $z$, 
we have that
\begin{equation} 
  \left(\prod_{x\in B_\dn^\prime}{b}_{x,\dn}\right)
  \left(\prod_{x\in B_\up^\prime}{b}_{x,\up}\right)
  \left(\prod_{x\in C_\dn}c_{x,\dn}\right)
  \left(\prod_{x\in C_\up}c_{x,\up}\right) 
    c_{z,\sigma}(c_{z,\up}^\dagger b_{z,\up} +
			     c_{z,\dn}^\dagger b_{z,\dn})\Phi=0
\end{equation}   
for arbitrary $B_\up^\prime, B_\dn^\prime\subset \La_2$ 
such that $\sum_{\sigma}(|B_\sigma^\prime|+|C_\sigma|)=\Ne-1$.
From this we obtain $f_2(B_\up^\prime\cup\{z\},B_\dn^\prime;C_\up,C_\dn)=0$
and $f_2(B_\up^\prime,B_\dn^\prime\cup\{z\};C_\up,C_\dn)=0$ 
for $z\notin C_\up\cup C_\dn$,
i.e., that $f_2(B_\up,B_\dn;C_\up,C_\dn)=0$ if
there exists a site $z$ such that $z\in B_\up\cup B_\dn$ and 
$z\notin C_\up\cup C_\dn$.
Therefore, we conclude that $f_2(B_\up,B_\dn;C_\up,C_\dn)=0$ 
unless all the conditions 
$C_\up\cap C_\dn=\emptyset$, $B_\up\cap B_\dn=\emptyset$ and $(B_\up\cup B_\dn)
\subset (C_\up\cup C_\dn)$
are satisfied.

Taking into account the above result and noting \eqref{eq:tildeb} and
\eqref{eq:inverse-tildeb},
we expand $\Phi$ as
\begin{equation}
 \Phi=\sum_{B_\up,B_\dn\subset\La_1}\sum_{C_\up,C_\dn\subset\La_1}
             \chi\left[\sum_{\sigma}|B_\sigma|\ge \sum_{\sigma}|C_\sigma|\right]
           f_1(B_\up,B_\dn;C_\up,C_\dn)\Phi_1(B_\up,B_\dn;C_\up,C_\dn),
\label{eq:Phi-expansion1}
\end{equation}
with coefficients $f_1$,
where we define $\Phi_1$ as in \eqref{eq:Phi2}, replacing subsets $C_\sigma$
and $B_\sigma$ of $\La_2$ with those of $\La_1$.
Then, repeating the same argument as above, we conclude that
$f_1(B_\up,B_\dn;C_\up,C_\dn)=0$
unless  $C_\up\cap C_\dn=\emptyset$, $B_\up\cap B_\dn=\emptyset$ 
and $(B_\up\cup B_\dn) = (C_\up\cup C_\dn)$.
We thus reach the desired expression \eqref{eq:Phi-expansion0} of a
zero-energy state $\Phi$.

We shall next show that the coefficients $f$ in \eqref{eq:Phi-expansion0}
satisfy 
\begin{equation}
 f(C;\bsigma,\btau)=f(C^\prime;\bsigma^\prime,\btau^\prime)
 \label{eq:f-condition}
\end{equation}
if $\sum_{x\in\La_1}(\sigma_x+\tau_x)=\sum_{x\in\La_1}(\sigma_x^\prime+\tau_x^\prime)$.

For $z\in\La_2$ and $C\subset\La_1$
a straightforward calculation yields 
\begin{eqnarray}
\lefteqn{
  c_{z,\dn}c_{z,\up}
                    \left( \prod_{x\in C}\tilde{b}_{x,\sigma_x}^\dagger \right)
	            \left(\prod_{x\in C}c_{x,\tau_x}^\dagger \right)
                                                                 \Phi_0
         }
           \nonumber\\
     &=&      
	     \sum_{y,y^\prime\in C} 
             \tilde{\varphi}_{y}(z)\tilde{\varphi}_{y^\prime}(z)\sgn[y,y^\prime;C]  
              \left(
		\chi[\sigma_{y}=\up,\sigma_{y^\prime}=\dn]
		-\chi[\sigma_{y}=\dn,\sigma_{y^\prime}=\up]
	       \right)
            \nonumber\\
      &&   \hspace*{3cm}
              \times
	       \left(\prod_{x\in C\backslash\{y,y^\prime\}}\tilde{b}_{x,\sigma_x}^\dagger\right)
	      \left(\prod_{x\in C}c_{x,\tau_x}^\dagger\right)
	       \Phi_0.
\label{eq:Phi-condition-U} 
\end{eqnarray}
Here 
$\sgn[\cdots]$ is a sign factor arising from exchanges of 
the fermion operators. 
By definition, for any $y,y^\prime$ in $C\subset\La_1$, 
$\tilde{\varphi}_y(z)\tilde{\varphi}_{y^\prime}(z)$ with $z\in\La_2$ is non-vanishing. 
We thus have from \eqref{eq:condition-U} and \eqref{eq:Phi-condition-U} that
$f(C;\bsigma,\btau)=f(C;\bsigma[y\leftrightarrow y^\prime],\btau)$, 
where $\bsigma[y\leftrightarrow y^\prime]$ is a spin configuration
obtained 
by switching $\sigma_{y}$ with $\sigma_{y^\prime}$.

It is also easy to see that
\begin{eqnarray}
\lefteqn{(c_{z,\up}^\dagger b_{z,\up}+c_{z,\dn}^\dagger b_{z,\dn})
	   \left( \prod_{x\in C}\tilde{b}_{x,\sigma_x}^\dagger \right)
           \left(\prod_{x\in C}c_{x,\tau_x}^\dagger \right)
            \Phi_0}
 \nonumber\\
&=&\sgn[z;C]\left(\chi[\sigma_z=\up,\tau_z=\dn]-\chi[\sigma_z=\dn,\tau_z=\up]\right)
 \nonumber\\
&& \hspace*{3cm}\times
           c_{z,\up}^\dagger c_{z,\dn}^\dagger
	   \left( \prod_{x\in C\backslash\{z\}}\tilde{b}_{x,\sigma_x}^\dagger \right)
           \left(\prod_{x\in C\backslash\{z\}}c_{x,\tau_x}^\dagger \right)
	   \Phi_0
\label{eq:Phi-condition-J}
\end{eqnarray}
for $z\in C\subset\La_1$. 
We thus have from \eqref{eq:condition-J} and \eqref{eq:Phi-condition-J}
that
$f(C;\bsigma,\btau)=f(C;\bsigma[z],\btau[z])$ where $\bsigma[z]$ and
$\btau[z]$ are obtained by switching $\sigma_z$ and $\tau_z$ in
$\bsigma$ and $\btau$. 

So far we have shown \eqref{eq:f-condition} with $C=C^\prime$.
We shall complete the proof of the claim by
examining~\eqref{eq:condition-J} with $z\in\La_2$.  

By using \eqref{eq:b} and \eqref{eq:inverse-tildeb},
\eqref{eq:condition-J} is rewritten as
\begin{equation}
  \sum_{y,y^\prime\in\La_1}\varphi_z(y)\varphi_z(y^\prime)
         (\tilde{b}_{y,\up}^\dagger c_{y^\prime,\up}+\tilde{b}_{y,\dn}^\dagger c_{y^\prime,\dn})\Phi=0.
\label{eq:condition-J-2}
\end{equation}
Let $y,y^\prime$ be a pair of sites with $|y-y^\prime|=1$ in $\La_1$.
For this pair, there exists a site $z$ in $\La_2$ such that
$\varphi_z(y)\varphi_z(y^\prime)\ne0$.
Let $D$ be an arbitrary subset of $\La_1$ such that $y,y^\prime\notin D$ and $|D|=\Ne/2-1$.
Then,  noting \eqref{eq:condition-J-2},
we obtain from
\begin{eqnarray}
\lefteqn{
  \left(\prod_{x\in D}c_{x,\upsilon_x}\right)
  \left(\prod_{x\in D\cup\{y,y^\prime\}}\tilde{b}_{x,\iota_x}\right)
     (c_{z,\up}^\dagger {b}_{z,\up}+c_{z,\dn}^\dagger {b}_{z,\dn})\Phi}\\
&=&
 \varphi_{z}(y)\varphi_z(y^\prime)
  \left(\prod_{x\in D}c_{x,\upsilon_x}\right)
  \left(\prod_{x\in D\cup\{y,y^\prime\}}\tilde{b}_{x,\iota_x}\right)
     (\tilde{b}_{y,\iota_y}^\dagger c_{y^\prime,\iota_y} 
      +\tilde{b}_{y^\prime,\iota_{y^\prime}}^\dagger c_{y,\iota_{y^\prime}})\Phi
=0
\end{eqnarray}
that
\begin{equation}
 f(D\cup\{y\};\bsigma,\btau)=f(D\cup\{y^\prime\};\bsigma^\prime,\btau^\prime)
\end{equation}
where $\sigma_y=\iota_{y^\prime}$, $\tau_y=\iota_{y}$,
$\sigma_{y^\prime}=\iota_y$, $\tau_{y^\prime}=\iota_{y^\prime}$ and
$\sigma_x=\sigma_x^\prime=\upsilon_x$,
$\tau_x=\tau_x^\prime=\iota_x$ for $x\in D$.
Let $C$ be an arbitrary subset of $\La_1$ with $|C|=\Ne/2$ 
and let $C_{y\rightarrow y^\prime}$ be a subset 
which is obtained by removing $y$ from and adding $y^\prime$ to $C$
(we assume $y\in C$ and $y^\prime\notin C$).
Then the above result implies that \eqref{eq:f-condition} holds for
$C^\prime=C_{y\rightarrow y^\prime}$ if $y$ and $y^\prime$ satisfy $|y-y^\prime|=1$.
It is easy to check that, for arbitrary subsets $C$ and $C^\prime$ of
$\La_1$ with $|C|=|C^\prime|$,
there exist pairs $(\{y_l,y_l^\prime\})_{l=1}^n$ with $|y_l-y_l^\prime|=1$
such that $C^1=C_{y_1\to y_1^\prime},C^2=C_{y_2\to y_2^\prime}^1,
\cdots,C^\prime=C_{y_n\to y_n^\prime}^{n-1}$. 
This completes the proof of the desired relation~\eqref{eq:f-condition}.

We write $f(M)$ for $f(C;\bsigma,\btau)$ with $\sum_{x\in\La_1}(\sigma_x+\tau_x)=M$.
By using \eqref{eq:f-condition}, \eqref{eq:Phi-expansion0} is rewritten
as
\begin{eqnarray}
 \Phi&=&\sumtwo{C\subset\La_1}{|C|=\Ne/2}\sum_{M=-\Ne/2}^{\Ne/2}
 \sum_{\bsigma,\btau}f(M)\chi[\sum_{x\in\La_1} (\sigma_x+\tau_x)=M]
                    \left(\prod_{x\in C}\tilde{b}_{x,\sigma_x}^\dagger\right)
                     \left(\prod_{x\in C}c_{x,\tau_x}^\dagger\right)
                    \Phi_0 \nonumber\\
     &=&\sumtwo{C\subset\La_1}{|C|=\Ne/2}
         \sum_{M=-\Ne/2}^{\Ne/2}
                       f^\prime(M)\left(S^{-}_{\mathrm{tot}}\right)^{\frac{\Ne}{2}-M}
                    \left(\prod_{x\in C}\tilde{b}_{x,\up}^\dagger\right)
                    \left(\prod_{x\in C}c_{x,\up}^\dagger\right)
                    \Phi_0 \nonumber\\
     &=& \sum_{M=-\Ne/2}^{\Ne/2}
                       f^{\prime\prime}(M)\left(S^{-}_{\mathrm{tot}}\right)^{\frac{\Ne}{2}-M}
                    \PhiG 
\end{eqnarray}
where 
\begin{eqnarray}
f^\prime(M) &=& \frac{(\Ne/2+M)!(\Ne/2-M)!}{\Ne!}f(M), \\
f^{\prime\prime}(M)&=&(-1)^{\frac{\Ne}{4}\left(\frac{\Ne}{2}-1\right)}
                      \frac{1}{(\Ne/2)!}f^\prime(M), 
\end{eqnarray}
and
the total spin lowering operator $S^{-}_{\mathrm{tot}}$ is
defined as $S^{-}_{\mathrm{tot}}=\sum_{x\in\La}c_{x,\dn}^\dagger
c_{x,\up}$.
To get the final expression we used
\begin{equation}
\label{eq:tzeta2}
 \tilde{\zeta}_{\up\up}=\sum_{x\in\La_1,y\in\La_2}G_{x,y}\tilde{b}_{x,\up}\tilde{b}_{y,\up}
                +\sum_{x\in\La_2,y\in\La_1}G_{x,y}\tilde{b}_{x,\up}\tilde{b}_{y,\up}
               =\sum_{x\in\La_1}c_{x,\up}\tilde{b}_{x,\up}.
\end{equation} 
This completes the proof of Proposition \ref{prop:ground-state}.
\bigskip\\
\textbf{Proof of Proposition \ref{prop:order}}.
As in the case of $\tilde{\zeta}_{\up\up}$, we have
\begin{equation}
\label{eq:zeta2}
 \zeta_{\up\up}=\sum_{x\in\La_1}b_{x,\up}c_{x,\up}.
\end{equation}
The momentum representation of $\zeta_{\up\up}$ is
\begin{equation}
 \zeta_{\up\up} = \sum_{k\in\calK} g_k \hat{c}_{(2,k),\up}\hat{c}_{(1,-k),\up},
 \label{eq:zeta3}
\end{equation}
where 
\begin{eqnarray}
\hat{c}_{(l,k),\sigma}&=&\frac{1}{\sqrt{L}}\sum_{x\in\La_l}\rme^{-\rmi kx}
                           c_{x,\sigma}, \\ 
\end{eqnarray}
with $l=1,2$ and
\begin{equation}
   \calK=\left\{ k = \frac{2\pi}{L}n~\Big{|}~-\frac{L-1}{2}\le n \le \frac{L-1}{2}
          \right\}.
\end{equation}

By using \eqref{eq:tzeta2} and \eqref{eq:zeta2}, we have
\begin{eqnarray}
 \PhiG &\propto&
        \sumtwo{C\subset\La_1}{|C|=\Ne/2}\left(\prod_{x\in C}
                                             \tilde{b}_{x,\up}^\dagger
                                                    c_{x,\up}^\dagger
                                         \right)\Phi_0
      \propto
        \sumtwo{C\subset\La_1}{|C|=\Ne/2}\left(\prod_{x\in C}
                                             \tilde{b}_{x,\up}^\dagger
                                                    c_{x,\up}^\dagger
                                         \right)
                                        \left(\prod_{x\in \La_1}
                                                   {b}_{x,\up}
                                                    c_{x,\up}\right)
                                        \PhiF
  \nonumber\\
    &\propto&
        \sumtwo{C\subset\La_1}{|C|=\Ne/2}
                                        \left(\prod_{x\in \La_1\backslash C}
                                                   {b}_{x,\up}
                                                    c_{x,\up}\right)
                                        \PhiF
      \propto \left(\zeta_{\up\up}\right)^{\frac{\Nh}{2}}\PhiF
\label{eq:PhiG-zeta}
\end{eqnarray}
where $\PhiF=\left(\prod_{x\in\La}c_{x,\up}^\dagger\right)\Phi_0$ and
$\Nh=|\La|-\Ne$, which is the number of holes counted from half-filling of electrons.
We use the above representation of $\PhiG$ with $\zeta_{\up\up}$ in
\eqref{eq:zeta3}
to estimate a lower bound on $\mu(\nu)$.

It is easy to check the anticommutation relation 
$\{\hat{c}_{(l,k),\sigma}^\dagger, \hat{c}_{(l^\prime,k^\prime),\sigma^\prime}\}
 =\delta_{l,l^\prime}\delta_{k,k^\prime}\delta_{\sigma,\sigma^\prime}$,
and by using this relation we have
\begin{equation}
   \zeta_{\up\up}^\dagger\zeta_{\up\up}
       = \sum_{k\in\calK}|g_k|^2 
         - \sum_{k\in\calK}\sum_{l=1,2} |g_k|^2\hat{c}_{(l,k),\up}\hat{c}_{(l,k),\up}^\dagger
         +\zeta_{\up\up}\zeta_{\up\up}^\dagger.
\end{equation}
We furthermore have
\begin{equation}
   \hat{c}_{(1,k),\up}\hat{c}_{(1,k),\up}^\dagger \zeta_{\up\up}
      =g_{-k}\hat{c}_{(2,-k),\up}\hat{c}_{(1,k),\up}
       +
         \zeta_{\up\up}\hat{c}_{(1,k),\up}\hat{c}_{(1,k),\up}^\dagger,
\end{equation}
which leads to
\begin{equation}
 \hat{c}_{(1,k),\up}\hat{c}_{(1,k),\up}^\dagger\left(\zeta_{\up\up}\right)^m\PhiF
 =mg_{-k} \hat{c}_{(2,-k),\up}\hat{c}_{(1,k),\up}\left(\zeta_{\up\up}\right)^{m-1}\PhiF
\end{equation}
for integers $m$ and thus
\begin{equation}
 \left(\zeta_{\up\up}\right)^{\frac{\Nh}{2}-m}
      \hat{c}_{(1,k),\up}\hat{c}_{(1,k),\up}^\dagger
 \left(\zeta_{\up\up}\right)^m\PhiF
 =\frac{2m}{\Nh}\hat{c}_{(1,k),\up}\hat{c}_{(1,k),\up}^\dagger
             \left(\zeta_{\up\up}\right)^{\frac{\Nh}{2}}\PhiF.
\end{equation}
The same relation holds for
$\hat{c}_{(2,k),\up}\hat{c}_{(2,k),\up}^\dagger$. 
We thus obtain
\begin{eqnarray}
\frac{\langle\PhiG, \Delta\Delta^\dagger \PhiG\rangle}
     {\langle\PhiG,\PhiG\rangle}
 &=& \frac{\Nh}{2}\frac{1}{L^2}\sum_{k\in\calK}|g_k|^2
  \nonumber\\
 && -\frac{1}{2}\left(\frac{\Nh}{2}-1\right)\frac{1}{L^2}
  \sum_{k\in\calK}\sum_{l=1,2}|g_k|^2
   \frac{\langle\PhiG, \hat{c}_{(l,k),\up} \hat{c}_{(l,k),\up}^\dagger \PhiG\rangle}
        {\langle\PhiG,\PhiG\rangle}
  \nonumber\\
 &\ge&
  \frac{\Nh}{4}\frac{1}{L^2}
  \sum_{k\in\calK}\sum_{l=1,2}|g_k|^2
   \frac{\langle\PhiG, \hat{c}_{(l,k),\up}^\dagger \hat{c}_{(l,k),\up} \PhiG\rangle}
        {\langle\PhiG,\PhiG\rangle}
  \nonumber\\
 & \ge &
    \frac{|\La|-\Ne}{2|\La|}\frac{2}{L}
  \sum_{m=1}^{\Ne/2}|g_{k(m)}|^2,
\end{eqnarray}
where we arranged the elements in $\calK$ as
$k(1),k(2),\dots,k(L)$
so that
$|g_{k(m)}|^2\le|g_{k(m^\prime)}|^2$ if $m\le m^\prime$.
Taking the limit $|\La|,\Ne\to\infty$ with $\Ne/(2|\La|)$ converging to
$\nu$ completes the proof.
\section{Remarks}
\label{s:remarks}
In this section we remark some
aspects of our model and its ground states.

The state $(S_\mathrm{tot}^{-})^{\frac{\Ne}{2}-M}\PhiG$ 
is the representative of the ground state in the subspace 
where the third component of the total spin is $M$.
If we use the expression~\eqref{eq:PhiG-zeta} of $\PhiG$, this becomes
$(\zeta_{\up\up})^{\frac{\Nh}{2}}\Phi_{M}$ where 
$\Phi_M=(S_\mathrm{tot}^{-})^{\frac{\Ne}{2}-M}\PhiF$.
This implies that the ground state can be regarded 
as a hole-condensation state in which all holes form the spin-triplet pair
in the back ground of the fullypolarized ferromagnetic state.

By using the so-called d vector $\vecd=(d_1,d_2,d_3)$, one can 
obtain a useful representation of a pair operator~\cite{Leggett75}.
In our case, we define
\begin{equation}
 \tilde{\zeta}_{\mbox{\small\boldmath $d$}}^\dagger = \sum_{x,y\in\La_1}
                                 \sum_{\sigma,\tau=\up\dn}
                                  F_{x,y}^{\sigma\tau}(\vecd) 
                                 \tilde{b}_{x,\sigma}^\dagger \tilde{b}_{y,\tau}^\dagger
\end{equation}
with
\begin{equation}
  \left(\begin{array}{@{\,}cc}
            F_{x,y}^{\up\up}(\vecd) & F_{x,y}^{\up\dn}(\vecd) \\
            F_{x,y}^{\dn\up}(\vecd) & F_{x,y}^{\dn\dn}(\vecd)
         \end{array} 
              \right)
 =G_{x,y}
    \left(\begin{array}{@{\,}cc}
            -d_1 + \rmi d_y & d_3 \\
             d_3            & d_1 + \rmi d_y
         \end{array} 
              \right).
\end{equation}
The pair operator $\tilde{\zeta}_{\up\up}^\dagger$ 
corresponds to
the case of the complex d vector $\vecd=(-1/2,-\rmi/2,0)$,  
and thus the ground state $\PhiG$ is a nonunitary spin-triplet pairing
state similar to the one describing the $A_1$-phase of superfluid
$^3\mathrm{He}$ in the magnetic field~\cite{Leggett75}.
It is noted that, unlike the case of $A_1$-phase realized in
the magnetic field, the states 
\begin{equation}
\left(\tilde{\zeta}_{\mbox{\small\boldmath $d$}}^\dagger\right)^{\frac{\Ne}{2}}\Phi_0
\end{equation}
with any $\vecd$ obtained by rotating $(-1/2,-\rmi/2,0)$ are the
ground states of $H$, since our Hamiltonian $H$ has spin-rotation symmetry.
The direction of the magnetic moment is given by the cross product
$\rmi \vecd \times \vecd^\ast$ where $\ast$ means the complex conjugation.

Let $\calB$ be the collection of pairs $\{x,y\}$ of sites such that
$x\in\La_1$, $y\in\La_2$ and $|x-y|=1/2$.
Then, we have from \eqref{eq:zeta2} that 
\begin{equation}
 \Delta^\dagger=\frac{1}{L}\sum_{\{x,y\}\in\calB}c_{x,\up}^\dagger c_{y,\up}^\dagger.
\end{equation} 
Therefore, the order parameter $\mu(\nu)$ measures 
the long-range correlation between local spin-1 electron pairs, and
Proposition~\ref{prop:order} implies that there exists this long-range correlation
in the ground state of $H$ in the thermodynamic limit. 

With respect to the ground state $\PhiG$ 
expectation values of operators constituted of either annihilation
or creation operators, 
such as $\Delta$ and $\hat{c}_{(2,k),\up}\hat{c}_{(1,-k),\up}$, 
are always zero
since there are exactly $\Ne$ electrons in $\PhiG$.
In order to obtain 
a particle number symmetry breaking ground state,
which is usually discussed in mean field approximations,
we need to form a linear combination of $\PhiG$ 
with different electron numbers.
Here let us consider the Bardeen-Cooper-Schrieffer type state
\begin{equation}
 \PhiG^\prime=\prod_{k\in\calK}\left(1+g_k \hat{c}_{(2,k),\up}\hat{c}_{(1,-k),\up}\right)
              \PhiF.
\end{equation}   
It is noted that the projection of $\PhiG^\prime$ onto the Hilbert space with
the fixed electron number is proportional to $\PhiG$ and thus
$\PhiG^\prime$ attains the ground state energy of $H$.
Then, for $\PhiG^\prime$ 
the expectation values of pair annihilation and creation operators
are calculated as
\begin{equation}
  \frac{\langle\PhiG^\prime,~\hat{c}_{(2,k),\up}\hat{c}_{(1,-k),\up}\PhiG^\prime\rangle}
       {\langle\PhiG^\prime,~\PhiG^\prime\rangle}
 =
 \frac{\langle\PhiG^\prime,~\hat{c}_{(1,-k),\up}^\dagger\hat{c}_{(2,k),\up}^\dagger\PhiG^\prime\rangle}
       {\langle\PhiG^\prime,~\PhiG^\prime\rangle}
 =\frac{g_k}{1+g_k^2},
\end{equation} 
which is finite for $-\pi<k<\pi$.

When we set $U=0$, the model with $2t=J$
has degenerate ground states and 
does not exhibit ferromagnetism.
In fact, states of the form
\begin{equation}
 \left(\tilde{\zeta}_{\dn\dn}^\dagger\right)^{N_{\dn\dn}}
 \left(\tilde{\zeta}_{\up\dn}^\dagger\right)^{N_{\up\dn}}
 \left(\tilde{\zeta}_{\up\up}^\dagger\right)^{N_{\up\up}}\Phi_0
\end{equation}
with non-negative integers $N_{\dn\dn}$, $N_{\up\dn}$ and $N_{\up\up}$
are zero-energy states of $\tHintJ$,
since we have 
\begin{equation}
(c_{x,\up}^\dagger b_{x,\up}+c_{x,\dn}^\dagger b_{x,\dn})\tilde{\zeta}_{\sigma\tau}^\dagger
=
\tilde{\zeta}_{\sigma\tau}^\dagger(c_{x,\up}^\dagger b_{x,\up}+c_{x,\dn}^\dagger b_{x,\dn})
\end{equation}
for any $x\in\La$ and $\sigma,\tau=\up,\dn$, as in
\eqref{eq:tzeta-commutation}.
The on-site repulsion removes this degeneracy and generates 
the unique ferromagnetic pairing ground state.

We can extent the present model, which has spin-rotation symmetry,
to anisotropic spin-interaction cases.
Consider the Hamiltonian 
\begin{equation}
 H^{\alpha,\beta} = \Hhop+\HintJ^{\alpha,\beta}+\HintU
\end{equation}
where $\HintJ^{\alpha,\beta}$ is given by
\begin{equation}
 \HintJ^{\alpha,\beta}
        =-J\sum_{x\in\La}\left\{
           \frac{n_x n_x^b}{4}
             +\beta S_{x,1}^b S_{x,1} + \beta S_{x,2}^b S_{x,2} 
             +(\alpha-\alpha\beta+\beta)S_{x,3}^b S_{x,3}          
          \right\}
 \label{eq:anisotropic-Hamiltonian}
\end{equation}
with $\alpha=\pm1$ and $\beta$ in $0<\beta\le1$.
It is noted that $H^{\pm1,1}$ becomes $H$.
 
The following results are obtained for the Hamiltonian $H^{\alpha,\beta}$.
In the case of $\alpha=1,~0<\beta<1,~2t=J,~U>0$, 
for even $\Ne$ in $2\le\Ne\le|\La|$,
there exist exactly two
ground states of $H^{\alpha,\beta}$, which are given by
\begin{equation}
 \PhiG^{\sigma\sigma} = \left(\tilde{\zeta}_{\sigma\sigma}^\dagger\right)^\frac{\Ne}{2}\Phi_0
\end{equation}
with $\sigma=\up,\dn$.       
In this case, therefore, the ground states of $H^{\alpha,\beta}$ exhibit ferromagnetism 
and condensation of the spin-1 electron pairs whose spins point in the
same direction 
parallel to the third component axis.
In the case of $\alpha=-1,~0<\beta<1,~2t=J,~U=0$, 
the ground state of $H^{\alpha,\beta}$ is unique 
for even $\Ne$ in $2\le\Ne\le2|\La|$, and is given by
\begin{equation}
 \PhiG^{\up\dn} = \left(\tilde{\zeta}_{\up\dn}^\dagger\right)^\frac{\Ne}{2}\Phi_0.
\end{equation}
In this case the ground state exhibits condensation of the spin-1
electron pairs whose spins point in the plane perpendicular to 
the third component axis, but does not exhibit ferromagnetism.
One can prove these results in the same way 
as in section~\ref{s:proof}, by noting the relation 
\begin{equation}
 H^{\alpha,\beta}=\beta\tHintJ
                        +(1-\beta)\tHintJ^{\alpha}+\HintU
\end{equation}
with
\begin{equation}
 \tHintJ^{\alpha}= \frac{J}{2}\sum_{x\in\La}\sum_{\sigma=\up,\dn}
                              b_{x,\sigma}^\dagger c_{x,\alpha\sigma} 
                              c_{x,\alpha\sigma}^\dagger b_{x,\sigma}.
\end{equation}

In this paper, we treat only the case of $2t=J$ where 
the exact spin-1 pairing ground
states can be constructed.
Apart from $2t=J$ we have the following exact results.
When we set $J=0$, the model becomes the decoupled two Hubbard chains, 
and thus the ground state of $H$ is non-magnetic.
In another limit of $U\to\infty$, where double occupancies at sites
are forbidden, 
it can be proved, by using the Perron-Frobenius theorem~\cite{comment3},
that the ground state of $H$ for $2t<J,~0<\Ne\le|\La|$ 
with open boundary conditions
is ferromagnetic.  
So we expect that the ground state phase diagram
of our model has a rich structure
depending on $J$, $U$, and the electron filling factor
$\nu$
(and also the anisotropy parameters $\alpha$ and $\beta$ in the
anisotropic cases).
It is desirable to describe the phase diagram in detail, 
but it is beyond of our scope at the present time.
We leave this problem for future study.   
\section{Higher Dimensional Cases}
%
Let $L$ be a positive odd integer and define $\La_1=[0,L-1]^d\cap
\mathbf{Z}^d$.
Let $\ba$ be a vector whose all components are $1/2$, and define
$\La_2=\{\bx+\ba~|~\bx\in\La_1\}$ and $\La=\La_1\cup\La_2$.
We impose periodic
boundary conditions in all directions on $\La$.
For each $\bx\in\La$, let $\bvarphi_\sbx=(\varphi_\sbx(\by))_{\sby\in\La}$ 
be a vector whose components are given
by
\begin{equation}
 \varphi_\sbx(\by)=\left\{
	       \begin{array}{@{\,}l}
		     1 ~~~\mbox{if $\displaystyle |\by-\bx|=|\ba|=\sqrt{{d}/{2^d}}$}\\
                     0 ~~~\mbox{otherwise}.   
	       \end{array}   
              \right.
\end{equation}
Recalling that the periodic boundary conditions are adopted, 
one notices that 
for each $\bx\in\La_1$, sites $\by\in\La_2$ can be written as
$\by=\ba+\bx+\sum_{l=1}^d n_l(\bx,\by)\bdelta_{l}$ where 
$\bdelta_l$ is the unit vector along the $l$-axis, and
$n_l(\bx,\by)$ with $l=1,\dots,d$ 
are integers in $0\le n_l(\bx,\by) \le L-1$.
Then, for $\bx\in\La_1$, let
$\btvarphi_\sbx=(\tilde{\varphi}_{\sbx}(\by))_{\sby\in\La}$ 
be a vector whose components are given by  
\begin{equation}
 \tilde{\varphi}_\sbx(\by)=\left\{
	       \begin{array}{@{\,}rl}
		     1/2^d &~~~\mbox{if $\by\in\La_2$ and $ \sum_{l=1}^d n_l(\bx,\by)$ is even}\\
                     -1/2^d &~~~\mbox{if $\by\in\La_2$ and $\sum_{l=1}^d n_l(\bx,\by)$ is odd}\\   
                     0 &~~~\mbox{if $\by\in\La_1$ },   
	       \end{array}   
              \right.
\end{equation}
and for $\bx\in\La_2$, let $\btvarphi_\sbx$ be a vector obtained by 
$\tilde{\varphi}_{\sbx}(\by)=\tilde{\varphi}_{\sbx-\sba}(\by-\ba)$.
Let $[G_{\sbx,\sby}]_{\sbx,\sby\in\La}$ be an antisymmetric matrix whose elements
are given by
\begin{equation}
 G_{\sbx,\sby}= \left\{
	       \begin{array}{@{\,}rl}
		     -1/2  &~~~\mbox{if $\bx\in\La_1$, $\by\in\La_2$ and $|\bx-\by|=|\ba|$}\\
                     1/2 &~~~\mbox{if $\bx\in\La_2$, $\by\in\La_1$ and $|\bx-\by|=|\ba|$}  \\   
                     0  &~~~\mbox{otherwise}.   
	       \end{array}   
              \right.
\end{equation}
 
By using $\bvarphi_\sbx$, $\btvarphi_\sbx$ and
$[G_{\sbx,\sby}]_{\sbx,\sby\in\La}$ 
introduced as above, we
define $b$- and $\tilde{b}$-operators and pair operators 
$\tilde{\zeta}_{\sigma\tau}$ and ${\zeta}_{\sigma\tau}$,
and then define the Hamiltonian $H$,
as in section~\ref{s:definition and main results}.
For this $d$-dimensional Hamiltonian we can obtain the same results as
in Proposition~\ref{prop:ground-state} and \ref{prop:order}, where
$I(\nu)$ is replaced with
\begin{equation}
 I_d(\nu)=2\left(
           \frac{1}{(2\pi)^d}\int_{|k_1|\le \pi}\cdots \int_{|k_d|\le \pi}
            \chi[|g_\sbk|^2\le \epsilon(\nu)]|g_\sbk|^2 
           ~\mathrm{d}k_1\cdots\mathrm{d}k_d
               \right).
\end{equation}
Here $g_\sbk=2^d\prod_{l=1}^d\cos(k_l/2)$ and
${\epsilon}(\nu)$ is determined by
\begin{equation}
 \nu=\frac{1}{2}\left(
           \frac{1}{(2\pi)^d}\int_{|k_1|\le \pi}\cdots \int_{|k_d|\le \pi}
            \chi[|g_\sbk|^2\le \epsilon(\nu)] 
           ~\mathrm{d}k_1\cdots\mathrm{d}k_d
               \right).
\end{equation}
We can also obtain the $d$-dimensional anisotropic spin-interaction model 
as in section~\ref{s:remarks}.
\bigskip\\
\textbf{\Large Acknowledgements}
\bigskip\\
I would like to thank Masanori Yamanaka for useful discussions in the
early stage of the study. 
This work is supported by Grant-in-Aid for Young Scientists (B)
18740243, from MEXT, Japan.


\begin{thebibliography}{99}
\bibitem{comment1}
	Here $|\cdot|$ means the absolute value.
	The same symbol $|X|$ is used to denote the number of elements in a
	set $X$.
\bibitem{comment2}
        We assume that the product is ordered according to the labels
	associated with the sites.
\bibitem{comment3}
       In Ref.~\cite{Tasaki98} one can find 
        the detailed explanation for applications of 
        the Perron-Frobenius theorem to ferromagneitsm in one-dimension.
\end{thebibliography}
\end{document}